\documentclass{phb-proc4-auth}

\usepackage{graphicx}
\usepackage{amssymb}

\begin{document}
\begin{frontmatter}


\journal{SCES '04}


\title{Quantum phase transitions out of the heavy Fermi liquid}

%
%
%
%
%
%
\author[TS]{T. Senthil, }
\author[SS]{Subir Sachdev\corauthref{1}, }
\author[MV]{Matthias Vojta}

%

\address[TS]{Department of Physics, Massachusetts Institute of
Technology, Cambridge MA 02139, USA.}

\address[SS]{Department of Physics, Yale University, P.O. Box
208120, New Haven CT 06520-8120, USA.}

\address[MV]{Institut f\"ur Theorie der Kondensierten
Materie, Universit\"at Karlsruhe, 76128 Karlsruhe, Germany.}

%
%
%
%


%
%
%
%

\corauth[1]{Email: subir.sachdev@yale.edu}


\begin{abstract}

We review recent work on the instability of the heavy Fermi liquid
state (FL) of the Kondo lattice towards a magnetic metal in which
the local moments are not part of the Fermi sea. Using insights
drawn from the theory of deconfined quantum criticality of
insulating antiferromagnets, we discuss the possibility of a
direct second order transition between the heavy Fermi liquid and
such a  magnetic metal. We suggest the presence of at least two
distinct diverging time scales -  the shorter one describes
fluctuations associated with the reconstruction of the Fermi
surface, while a longer one describes fluctuations of the magnetic
order parameter. The intermediate time scale physics on the
magnetic side is suggested to be that of a novel fractionalized
Fermi liquid (FL*) state with deconfined neutral $S=1/2$
excitations. This could ultimately devolve into the magnetic phase
with conventional order at one of the larger time scales.
Experimental implications for this scenario are noted.


\end{abstract}

%
%



\end{frontmatter}

%
%
%
%
%

\section{Introduction}
\label{sec:intro}

A remarkable realization of Landau's Fermi liquid theory of metals
occurs in the `heavy fermions' class of materials. Indeed, it has
long been known that at low temperatures these can typically be
described within the Fermi liquid paradigm, albeit with strongly
renormalized parameters (for instance, effective masses of order
$10^2 - 10^3$ the band mass). Microscopically, this class of
material may be modelled as possessing localized magnetic moments
coupled to a separate set of conduction electrons. The heavy Fermi
liquid (see Section \ref{sec:hfl} below) is understood by a
lattice analog of the Kondo screening of the localized moments.
Crudely speaking, the localized moments ``dissolve'' into the
Fermi sea. Luttinger's theorem for the volume of the resulting
Fermi surface includes both the conduction electrons and those
forming the local moments.

In contrast, recent experimental work has focused on situations
where the Fermi liquid description of such metals appears to break
down in a rather strong manner. Such non-Fermi liquid behavior
occurs most strikingly in the vicinity of zero temperature phase
transitions out of the heavy Fermi liquid - typically to a
magnetically ordered metal. However there is little theoretical
understanding of this breakdown of the Fermi liquid paradigm.

Quite generally, two distinct kinds of magnetically ordered metals
are possible in heavy fermion materials. In one the magnetism is
to be viewed as a spin density wave instability arising out of the
parent heavy Fermi liquid state. Crudely speaking, this magnetism
may be viewed as arising from imperfectly Kondo-screened local
moments. We will refer to such a state as the SDW metal. A
different kind of magnetic metallic state is also possible where
the localized moments order due to RKKY exchange interactions, and
do not participate in the Fermi surface of the metal. The ``Kondo
order'' present in the heavy Fermi liquid is absent in such a
material. We will denote this second magnetically ordered metal as
a `local moment magnetic metal' or LMM metal. Often the
distinction between these two kinds of magnetic states can be made
sharply: the Fermi surfaces in the two states may have different
topologies (albeit the same volume modulo that of the Brillouin
zone of the ordered state) so that they cannot be smoothly
connected to one another.

There is a well developed theory for the quantum transition
between the heavy Fermi liquid and the SDW metal \cite{hertz}
which, however, often fails to produce the non-Fermi liquid
physics observed in experiments. It is therefore tempting to
assume that when non-Fermi liquid physics is seen at a magnetic
ordering transition out of the heavy Fermi liquid, the resulting
magnetic state is the LMM metal rather than the SDW metal.
However, there is no theoretical understanding of such a
transition. Indeed, a number of basic conceptual questions arise.
Can there be a second order transition where the `Kondo order' of
the heavy Fermi liquid disappears concomitantly with the
appearance of magnetic long range order? What is the theoretical
description of such a transition? Will it reproduce the observed
non-Fermi liquid physics in the heavy fermion metals near their
magnetic ordering transition?

The answers to these questions are not known with confidence at
present. In this paper we will review our recent work on related
questions in a number of simpler contexts. Based on the lessons
from these studies we will present some ideas on the transition
from the heavy Fermi liquid to the LMM metal, and their
implications for experiments.

\section{The heavy Fermi Liquid, FL}
\label{sec:hfl}
Much of our understanding of the heavy fermion compounds is based
on the Kondo lattice model:
\begin{equation}
  \mathcal{H}_K = \sum_k \epsilon_k c^{\dagger}_{k\alpha} c_{k\alpha} +
  \frac{J_K}{2}\sum_r
  \vec S_r \cdot c^{\dagger}_{r\alpha} \vec \sigma_{\alpha\alpha'}
  c_{r\alpha'}.
\label{KL}
\end{equation}
This model consists of a density $n_c$ of conduction electrons
$c_{k \alpha}$ with dispersion $\epsilon_k$ ($k$ is momentum and
$\alpha=\uparrow,\downarrow$ is a spin index) interacting with $f$
electron spins $\vec{S}_r$ ($r$ is a lattice position, and
$\vec{\sigma}$ are the Pauli matrices) via an antiferromagnetic
Kondo exchange coupling $J_K$.

\begin{figure}
\centering
\includegraphics[width=3.1in]{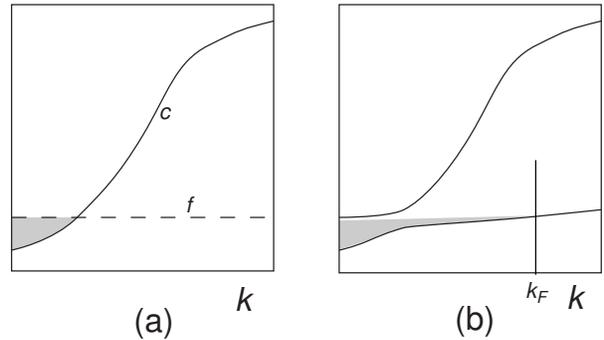}
\caption{\label{flatf} Conventional theory of the FL state. A
completely flat $f$ electron ``band'' (dashed line in (a)) mixes
with the conduction electrons to obtain the renormalized bands in
(b). There is a single Fermi surface at $k_F$ in the FL state
containing states whose number equals that of the $f$ and $c$
electrons combined.}
\end{figure}

There is a well accepted theory of the formation of a heavy
fermion liquid state (hereafter referred to as the FL state) in
this model \cite{cmv,aa,ajm}. The charge of the $f_{r \alpha}$
electrons is fully localized on the rare-earth sites, and so one
initially imagines that these electrons occupy a flat
dispersionless band, as shown in Fig~\ref{flatf}a.
This band has to be half-filled, and so we must place it at the
Fermi level. The $c_{r\alpha}$ electrons are imagined to occupy
their own conduction band. Next, it is argued \cite{cmv}, that the
Kondo exchange is equivalent to turning on a small hybridization
between these two bands. We represent this hybridization by a
non-zero expectation value of the bosonic operator
\begin{equation}
b_r \sim \sum_{\alpha} c^{\dagger}_{r\alpha} f_{r\alpha}.
\label{defb}
\end{equation}
With $\langle b_r \rangle$ non-zero, the two bands will mix and
lead to the renormalized bands shown in Fig~\ref{flatf}b. A
crucial point is that because the $f$ band was initially
dispersionless, there is absolutely no overlap between the
renormalized bands. Hence the occupied states are entirely within
the lower band, and one obtains a single Fermi surface within
wavevector $k_F$: the volume within $k_F$ is determined by the
total density of $c$ {\em and\/} $f$ electrons. This is precisely
the Fermi volume predicted in the limit of weak interactions
between the electrons by the Luttinger theorem. Moreover, as in
Fig~\ref{flatf}b, the Fermi surface is in a region where the
electrons are primarily have a $f$ character, and consequently the
band is still quite flat---this accounts for the large effective
mass of the fermionic quasiparticles.

A key technical feature of the theory of the FL state is presence of an
emergent compact U(1) gauge field. This is a consequence of the
quenching of the charge fluctuations on the $f$ electron sites
{\em i.e.\/} the constraint
\begin{equation}
\sum_\alpha f_{r \alpha}^{\dagger} f_{r \alpha}^{\vphantom
\dagger} = 1 \label{constraint1}
\end{equation}
is obeyed at every rare earth site. This constraint implies that
that theory is invariant under the spacetime dependent U(1) gauge
transformation
\begin{equation}
f_{r \alpha} \rightarrow f_{r \alpha} e^{i \phi_r (\tau)}
\label{gauge1}
\end{equation}
where $\tau$ is imaginary time. After integrating out high energy
degrees of freedom, this invariance leads to the emergence of a
dynamical compact U(1) gauge field $A_\mu$ ($\mu$ is a spacetime
index, and `compact' refers to the invariance of the theory under
$A_\mu \rightarrow A_\mu + 2 \pi$. Notice also that the bosonic
field $b$ in Eq.~(\ref{defb}) also carries a U(1) gauge charge.
The $b$ field is condensed in the FL state, and so in this state
the U(1) gauge theory can be considered to be in a `Higgs' phase.
The appearance of the $b$ Higgs condensate also means that $A_\mu$
fluctuations are quenched, and so are relatively innocuous in the
FL state.

We are interested here in the manner in which this heavy FL state
may be destroyed by perturbations at zero temperature ($T$). An
important early proposal made by Doniach \cite{doniach} was that
the state could be unstable to magnetic ordering of the $f$
moments, induced by a RKKY exchange coupling between them:
\begin{equation}
\mathcal{H}_H =  \sum_{rr'}J_H (r,r') \vec
  S_r \cdot \vec S_{r'} \,.
\label{KH}
\end{equation}
Assuming this metallic state is the SDW metal noted in
Section~\ref{sec:intro}, such a quantum phase transition can be
analyzed \cite{hertz} in a manner similar to the SDW instability
in an ordinary Fermi liquid. In such an approach one assumes that
once the heavy FL state has formed between the $f$ and $c$
electrons, the resulting quasiparticles lose memory of their
origin, and behave like ordinary `light' quasiparticles.

As discussed in Section~\ref{sec:intro}, here we will review our
recent work \cite{ffl,dcq} exploring another route to the
breakdown of the FL state of $\mathcal{H}_K + \mathcal{H}_H$.
Here, the Kondo lattice origins of the FL state play a central
role, and the focus is on the breakdown of the `hybridization' or
`Kondo screening' between the $f$ and $c$ bands. Magnetic order
may well appear at very low energies (in a LMM metal) once the FL
state has been disrupted; however, this will be viewed as a
secondary or `epiphenomenon', and the primary physics is that of
the destruction of the Higgs phase of the U(1) gauge theory.

Other distinct points of view are in Refs.~\cite{piers1,si,pepin}.

\section{The fractionalized Fermi liquid, FL*}
\label{sec:ffl}

\begin{figure}
\centering
\includegraphics[width=3.1in]{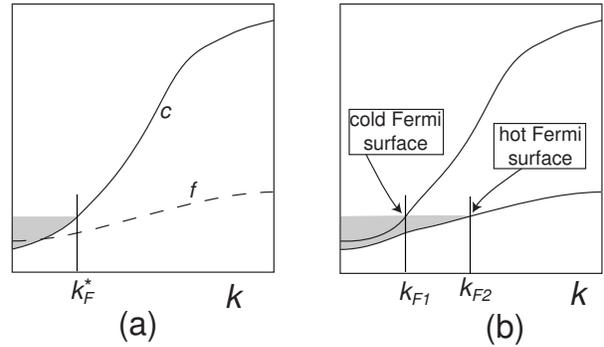}
\caption{\label{hotcold}(a) The FL* state with a cold Fermi
surface at $k_{F}^{\ast}$. The $f$ fermions form a spin liquid,
and are schematically represented by the dashed dispersion. (b)
The FL state which obeys the conventional Luttinger theorem, but
has two Fermi surfaces.}
\end{figure}

As noted above, the only active degrees of freedom on the $f$
sites are the spins, and this is captured by the $f_{r \alpha}$
operators. The exchange coupling in $\mathcal{H}_H$ can move this
spin between sites, and so the initial assumption above of a
dispersionless $f$ band may be questioned. So let us reconsider
the above argument starting from a $f$ band which has a dispersion
of order $J_H$, as shown in Fig~\ref{hotcold}(a).
When the hybridization energy, in Eq.~(\ref{defb}), is on a scale
larger than $J_H$, the physics is as in Section~\ref{sec:hfl}.
However, as the hybridization decreases, it is clear that
eventually there will be band overlap, and the FL state will have
2 occupied bands, as shown in Fig~\ref{hotcold}b, with two Fermi
wavevectors, $k_{F1}$ and $k_{F2}$. Nevertheless, this is still a
conventional FL state, in that the total number of states
contained between both Fermi surfaces still equals the total
number of $c$ and $f$ electrons, and the conventional
weak-coupling Luttinger theorem is still obeyed. However, it has
been argued \cite{bgg,ffl} that it is now possible to reach a
finite coupling quantum critical point where the Higgs condensate
disappears and $\langle b \rangle = 0$. The finite dispersion of
the $f$ band means that it is not always energetically preferable
to have a $b$ condensate, as was the case in Refs.~\cite{aa,ajm}.
Furthermore, careful gauge-theoretic arguments can be made
\cite{ffl} that the transition from a state with $\langle b
\rangle \neq 0$ to a state with $\langle b \rangle =0$ is indeed a
sharp phase transition at $T=0$. However, there is no conventional
Landau order parameter for this transition, and there is no analog
of such a transition at $T>0$ (although a $T>0$ transition between
a state with $\langle b \rangle \neq 0$ and a state with $\langle
b \rangle =0$ does appear in mean field theory \cite{coleman}).
Rather, it is a transition characterized by a change in the
`topological' character of the ground state wavefunction, in that
the fate of the $A_\mu$ field dynamics undergoes a qualitative
change.

So what is the nature of the state, which we denoted FL*, with
$\langle b \rangle = 0$ obtained by increasing $J_H$ ? First, it
is a Fermi liquid in the sense that the $c$ electrons form a
conventional Fermi surface, with sharp electron-like
quasiparticles at $k_F^\ast$ (see Fig~\ref{hotcold}a). However,
the volume enclosed by this Fermi surface contains states whose
number equals only the number of $c$ electrons; this violates the
conventional Luttinger theorem. Closely linked with this
non-Luttinger Fermi volume is the fate of the spin moments on the
$f$ electrons. These form a `spin liquid' state, which does not
break any lattice translational symmetries and fully preserves
spin rotation invariance ({\em i.e.\/} there is no magnetic
moment). The spin liquid state contains $S=0$ excitations of the
gauge field, $A_\mu$, which remain gapless if the U(1) gauge group
remains unbroken (spin liquid states with a $Z_2$ gauge group are
also possible), along with the now charge neutral $S=1/2$
$f_\alpha$ fermions. A fairly rigorous argument can be made
\cite{ffl,ashvin} showing that the unusual Fermi volume of the FL*
phase requires the emergent collective gauge excitations
represented by $A_\mu$.

Our work also presented \cite{ffl} a  description of the quantum
critical point between the FL and FL* phases. Within the FL phase,
as one moves towards the FL* phase, the two Fermi surfaces at
$k_{F1}$ and $k_{F2}$ behave differently. The Fermi surface at
$k_{F1}$ evolves into the Fermi surface at $k_{F*}$ in the FL*
state, and this remains `cold' near the quantum critical point:
there is no strong scattering of these quasiparticles, and the
electron quasiparticle residue remains finite across the
transition. In contrast, the Fermi surface at $k_{F2}$ becomes
`hot': the lifetime of the quasiparticles becomes short, and the
quasiparticle residue decreases and ultimately vanishes at the
quantum critical point. The dynamics of the quasiparticles
exhibits non-Fermi liquid behavior at the quantum critical point.
Note that this anomalous behavior extends across the entire hot
Fermi surface, in contrast to the isolated `hot spots' that appear
in the SDW theory \cite{hertz}. The self energies also have a full
momentum dependence, in contrast to theories of `local'
criticality \cite{si}.

\section{Deconfined quantum criticality in insulating antiferromagnets}
\label{sec:dqc}

It is useful to interpret the approach of the FL state to the FL*
state described above as a approach to a deconfinement transition
of the $f_\alpha$ fermions. In the FL state these are `bound' to
the $c_\alpha$ by the Higgs condensate of (\ref{defb}), the
ultimate low energy quasiparticles carry charge $-e$ and $S=1/2$,
and there are no strong $A_\mu$ fluctuations. At the quantum
critical point, neutral $f_\alpha$ quanta are liberated, and the
$A_\mu$ fluctuations become much stronger. The present theory
\cite{ffl} of this critical point treats these gauge fluctuations
perturbatively - for reasons explained in Ref. \cite{ffl} this may
be legitimate in spatial dimension $d = 3$, but not necessarily so
in $d =2$. In the latter case it is entirely possible that there
are additional non-perturbative effects which have not been fully
accounted for.

To explore the possible consequences of such deconfinement-driven
quantum criticality, we consider here a simpler insulating system
for which much progress has recently been made \cite{dcq} in
understanding the non-perturbative consequences of gauge
fluctuations. We describe the square lattice $S=1/2$ Heisenberg
antiferromagnet, important in its applications to the physics of
the cuprates. This is a model in the class $\mathcal{H}_H$ of the
$f$ moments alone, with no metallic charge carriers. When the
exchange interactions are predominantly nearest neighbor, the
ground state long-range N\'eel order, in which the moments are
polarized along opposite collinear directions on the two
sublattices. As is conventional, we characterize the local
orientation of this order by a unit vector field $\vec{n}_r
\propto (-1)^{x+y} \vec{S}_r$. It turns out to be useful to
further express the vector $\vec{n}_r$ in spinor variables by
\begin{equation}
\vec{n}_r = z_{r \alpha}^{\ast} \vec \sigma_{\alpha\beta} z_\beta
\label{cp1}
\end{equation}
where $z_{r \uparrow}$, $z_{r \downarrow}$ are complex spinors
obeying
\begin{equation}
\left| z_{r \uparrow} \right|^2 + \left| z_{r \downarrow}
\right|^2 =1 \;.\label{constraint2}
\end{equation}
As in Eqs.~(\ref{constraint1}), (\ref{gauge1}),
Eq.~(\ref{constraint2}) implies that the theory for the $z_\alpha$
has a compact U(1) gauge invariance
\begin{equation}
z_{r \alpha} \rightarrow z_{r \alpha} e^{i \phi_r (\tau)} \;,
\label{gauge2}
\end{equation}
and an associated compact U(1) gauge field $A_\mu$. In the present
situation, it is possible to use spin and lattice symmetries, and
the absence of gapless Fermi surfaces, to write down an explicit
effective action for the gauge theory in the continuum limit:
\begin{equation}
\mathcal{S}_z = \int d^2 r d\tau \Bigl[ |( \partial_\mu - i A_\mu)
z_{\alpha} |^2 + s |z_\alpha |^2 + u ( |z_\alpha |^2 )^2 \Bigr]
\label{sz}
\end{equation}
Here $s$ is a tuning parameter which we use to destroy N\'eel
order, and we will then describe the consequences of gauge
fluctuations at the resulting quantum critical point. The N\'eel
phase is obtained when $s<0$, and the $z_\alpha$ are condensed
with $\langle z_\alpha \rangle \neq 0$. In other words, the N\'eel
phase is the Higgs phase of the present gauge theory, in close
analogy to our discussion in Sections~\ref{sec:hfl},
\ref{sec:ffl} for the FL phase. Here too the Higgs condensate
quenches the gauge fluctuations, and morphs them into a $S=0$
collective mode of pairs of spin waves.

Now let us approach the quantum critical point, which is at
$s=s_c$ (say). Here, the $z_\alpha$ are gapless $S=1/2$ quanta
which interact via exchange of the gapless U(1) gauge quanta of
$A_\mu$. In a sense, the $z_\alpha$ quanta have been deconfined,
but some care has to used with this terminology because the
$z_\alpha$ also acquire an anomalous dimension \cite{sendai}.

We can continue the same analysis into the paramagnetic phase with
$s>s_c$ and $\langle z_\alpha \rangle =0$, where the properties of
the theory $\mathcal{S}_z$ are really quite simple, and describe a
U(1) spin liquid ground state. The $z_\alpha$ quanta are sharply
defined $S=1/2$ quasiparticles (spinons) and they interact via
exchange of $A_\mu$ quanta, which now have a true Maxwell-photonic
form.

\begin{figure}
\centering
\includegraphics[width=2.7in]{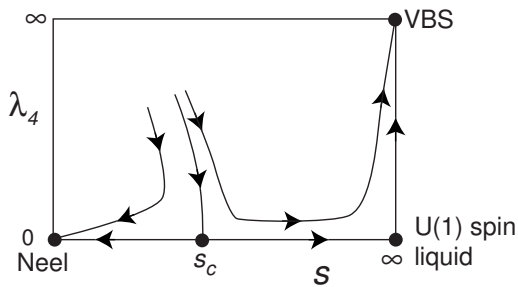}
\caption{\label{rgflow} Renormalization group flows for the
$S=1/2$ square lattice antiferromagnet. The transition between the
N\'{e}el and paramagnetic states is tuned by $s$, and $\lambda_4$
is a monopole fugacity whose bare value is generically nonzero.}
\end{figure}

While simple and direct, the above story turns out to be
fundamentally incomplete, and this breakdown likely has
implications for the FL to FL* transition we described earlier. In
taking the continuum limit to Eq.~(\ref{sz}), we have lost
information on the compactness of $A_\mu$ ({\em i.e.\/} invariance
under $A_\mu \rightarrow A_\mu + 2 \pi$). A compact U(1) gauge
theory in 2+1 dimensions allows {\sl monopole} point defects,
which are tunnelling events between sectors whose total flux
differs by $2 \pi$. In the language of the $\vec{n}$ field, these
are `hedgehog' defects. Furthermore, a lattice scale analysis of
the action of these monopoles shows that they carry Berry phases.
We refer the reader to Ref.~\cite{sendai} for a review of the
physics of these monopoles at and near the quantum critical point,
and merely summarize the results of such an analysis here. As
shown in Fig~\ref{rgflow}, it is useful to consider a
two-dimensional renormalization group (RG) flow in the tuning
parameter, $s$, and a certain quadrupled monopole fugacity,
$\lambda_4$.

The properties of the theory $\mathcal{S}_z$ are described by the
line $\lambda_4=0$. There is an unstable fixed point at $s=s_c$,
separating a flow towards the magnetically ordered N\'eel state
for $s<s_c$, from a $s>s_c$ flow towards the paramagnetic spin
liquid with a gapless U(1) photon and gapped spinon excitations.
The relevant RG eigenvalue at $s=s_c$ determines a spin
correlation length $\xi_{\rm spin} \sim |s-s_c|^{-\nu}$ which
diverges as the critical point is approached. For $s>s_c$ there is
an energy gap towards spinful excitations $\sim \xi_{\rm
spin}^{-1}$ which vanishes as $s \to s_c^+$.

Now consider a nonzero $\lambda_4$. Notice that $\lambda_4$ is
{\sl irrelevant} at the $s=s_c$, $\lambda_4=0$ fixed point. This
means that the above description of the critical properties of
$\mathcal{S}_z$ applies also to the underlying antiferromagnet.
There is, however, a crucial caveat. While $\lambda_4$ is
irrelevant at the quantum critical point, for $s>s_c$ the flow of
$\lambda_4$ eventually turns around and is attracted towards large
values of $\lambda_4$. This happens because $\lambda_4$ is a {\sl
relevant} perturbation to the large $s$ fixed point describing the
U(1) spin liquid phase. This phenomenon characterizes $\lambda_4$
as a {\sl dangerously irrelevant} coupling at the critical point.
For $s$ just above $s_c$, the value of $\lambda_4$ becomes very
small at length scales of order $\xi_{\rm spin}$ (or energy scales
of order $\sim \xi_{\rm spin}^{-1}$), but eventually becomes of
order unity or larger at a second length scale which we denote
$\xi_{\rm VBS}$. This behaves like $\xi_{\rm VBS} \sim \xi_{\rm
spin}^{\lambda}$, where $\lambda > 1$ is a critical exponent; so
we have $\xi_{\rm VBS} \gg \xi_{\rm spin}$. There is also a
corresponding energy scale $\sim \xi_{\rm VBS}^{-1}$ which is much
smaller than the energy gap to spin excitations.

\begin{figure}
\centering
\includegraphics[width=3.1in]{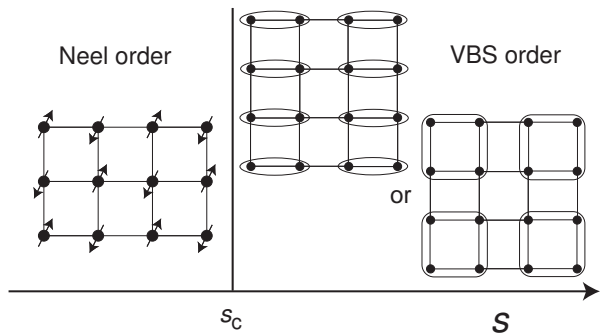}
\caption{\label{vbs} Phase diagram of the $S=1/2$ square lattice
antiferromagnet. The N\'eel state breaks spin rotation invariance.
The VBS state preserves spin rotation invariance but breaks
lattice rotation and translation symmetries. The ovals around the
lattice sites represent $S=0$ composites of the electrons on the
surrounded sites: these are meant to schematically indicated the
patttern of lattice symmetry breaking in the ground state
wavefunction. The VBS order for $s>s_c$ appears only at a
length/time scale which is much larger than the spin correlation
length/time $\sim \xi_{\rm spin}$.}
\end{figure}

What happens at the scale $\xi_{\rm VBS}$ ? In short, the
properties of the paramagnetic phase change completely. The
$S=1/2$ $z_\alpha$ quanta experience a confining force, and the
ground state breaks lattice translation and rotation symmetries by
the development of valence bond solid (VBS) order, as illustrated
in Fig~\ref{vbs}.
However, the remarkable fact is that all these changes to the
paramagnetic phase occur without modifying our earlier theory of
the quantum critical point. This is possible because the quantum
critical point has two diverging length/time scales, one much
larger than the other.

Returning to our discussion of the destruction of N\'eel order by
liberation of the $z_\alpha$ quanta, we see that this
`deconfinement' happens only at the $s=s_c$ quantum critical
point. The $z_\alpha$  are ultimately confined for all $s>s_c$,
but only at length/time scale so large that the confinement
physics has no effect on the critical theory.

\section{Implications for heavy electron criticality}
\label{sec:imp}

We now return to our discussion to the instability of the heavy
Fermi liquid. The analogy with our discussion in
Section~\ref{sec:dqc} should now be evident. In the insulating
magnets there is a direct second order transition between the
N\'eel state and a state with a very different kind of order (the
valence bond solid, VBS). This may be viewed as being analogous to
a direct second order transition between the heavy Fermi liquid
and the LMM metal where the loss of `Kondo order' happens
concomitantly with the appearance of magnetic order. Indeed, both
the N\'eel state (of the insulating magnet) and the FL state (of
the Kondo lattice)  are stable Higgs phases of a compact U(1)
gauge theory. They are unstable to a `deconfinement' transition to
the U(1) spin liquid and the FL* phases respectively. However,
such deconfined phases are rather fragile states of matter, and
can ultimately be unstable to confined phases with conventional
order. We reviewed above the instability of the U(1) spin liquid
into a VBS state. It is then natural to explore the instability of
the FL* state to the LMM, but in a manner that the quantum
criticality remains `deconfined'. In the insulating magnets, the
separation between the two competing orders (Neel and VBS) occurs
not as a function of a tuning parameter, but dynamically as a
function of length/time scale. Indeed, in the paramagnet, the loss
of Neel correlations occurs on one length scale $\xi$ while the
pinning of the VBS order appears on a much longer length scale
$\xi_{\rm VBS}$ that diverges as a power of $\xi$. By analogy, for
the heavy electron systems this strongly suggests that a direct
second order transition between the two appropriate competing
orders (Kondo order in FL and magnetic order in the LMM metal) is
possible but requires at least two diverging length/time scales at
the quantum critical point, with deconfinement evident only at the
shorter scale(s).

\begin{figure}
\centering
\includegraphics[width=2.7in]{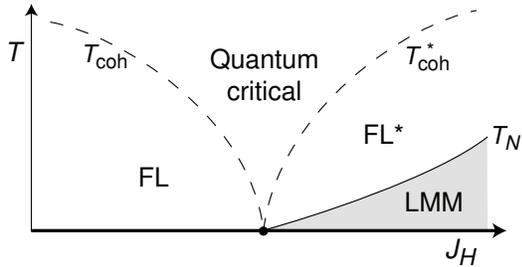}
\caption{\label{phasediag} Phase diagram near the conjectured FL
to LMM quantum transition. The two distinct energy scales are
manifested by the differing exponents by which $T_N$ and $T_{\rm
coh}$ (or $T_{\rm coh}^{\ast}$) approach the quantum critical
point.}
\end{figure}

These ideas suggest interesting and important directions for
experimental work on heavy electron systems. Are there indeed two
or more distinct diverging time/length scales near heavy fermion
critical points? In particular, we might expect that the
reconstruction of the Fermi surface happens (if it does so at all)
at a time scale which diverges slower than the time scale
associated with magnetic fluctuations. An immediate consequence is
that the Neel temperature, $T_N$, at which magnetism appears in
the LMM state, will vanish faster ({\em i.e.\/} with a larger
power of the tuning parameter across the quantum phase transition)
than the temperature scale, $T_{\rm coh}$, at which well-defined
quasiparticles appear at the large Fermi surface on the FL side
(see Fig~\ref{phasediag}).
Further, on the LMM side, there should be an intermediate
temperature regime $T_N \ll T \ll T^{*}_{\rm coh}$ in which the
properties are those of the small Fermi surface state FL$^*$
described in Section \ref{sec:ffl}. The two temperature scales
$T_{\rm coh}$ and $T^*_{\rm coh}$ must vanish identically on
approaching the quantum critical point from opposite sides.

More generally, the presence of two or more diverging length
scales will affect the scaling properties of a number of physical
quantities near the quantum critical point - probes of the
magnetic fluctuations will scale with a different length/time
scale from probes of fluctuations associated with the Fermi
surface structure. Experiments that elucidate the character of
these two possibly different kinds of fluctuations are important,
and would be extremely helpful in clarifying the physics behind
the non-Fermi liquid behavior near heavy fermion quantum critical
points.

This research was supported by the National Science Foundation
under grants DMR-0308945 (T.S.), DMR-0098226 (S.S.) and
DMR-0210790, PHY-9907949 at the Kavli Institute for Theoretical
Physics, and the DFG Center for Functional Nano\-structures
Karls\-ruhe (M.V.). T.S. and S.S. thank the Aspen Center for
Physics for hospitality. We would also like to acknowledge funding
from the NEC Corporation (T.S.), a John Simon Guggenheim
Foundation fellowship (S.S.), and an award from The Research
Corporation (T.S.).

%
%
%
%


\end{document}